\baselineskip 18pt

\def\a{\alpha}
\def\b{\beta}
\def\g{\gamma}
\def\t{\theta}
\def\P{{\bf P}}
\def\R{{\bf R}}
\def\p{{\bf p}}
\def\r{{\bf r}}
\def\J{{\bf J}}
\def\L{{\bf L}}
\def\M{{\bf M}}
\def\K{{\bf K}}
\def\ra{\rangle}
\def\la{\langle}
\def\ab{{\bf a}}

\noindent
{\bf MANY-BODY WIGNER QUANTUM SYSTEMS}

\vskip 32pt
\noindent
T. D. Palev\footnote*{Permanent address: Institute for Nuclear
Research and Nuclear Energy, 1784 Sofia, Bulgaria; E-mail:
palev@bgearn.acad.bg, stoilova@inrne.acad.bg}
and N. I. Stoilova*

\noindent
International Centre for Theoretical Physics, 34100 Trieste, Italy
\vskip 12pt

\vskip 48pt

{\bf Abstract.} We present examples of many-body Wigner quantum
systems. The position and the momentum operators $\R_A$ and
$\P_A,\; A=1,\ldots,n+1$, of the particles are noncanonical and
are chosen so that the Heisenberg and the Hamiltonian equations
are identical. The spectrum of the energy with respect to the
centre of mass is equidistant and has finite number of energy
levels. The composite system is spread in a small volume
around the centre of mass and within it the geometry is
noncommutative. The underlying statistics is an exclusion
statistics.

\vskip 48pt

\noindent
{\bf 1. Introduction}
\bigskip

\noindent
In the present paper we continue the study of the  Wigner quantum
systems (WQSs), initiated in [1-4]. Our interest in the subject
is stimulated from the observation that some WQSs show atractive
features, which cannot be achieved in the frame of the canonical
quantum mechanics.  The Wigner quantum system (WQS) from [1] (two
noncanonical, nonrelativistic point particles interacting via
harmonic potential), for instance, exhibits a quark like
structure: the composite system has finite  size,
both constituents are bound to each other; moreover the
geometry is noncommutative, the different coordinates do not
commute.  Another example following from [3,4]: two spinless
particles, curling around each other, produce an orbital
(internal angular) momentum $1/2$.

Here we extand the results of [1] to the case of any number of
particles. For definiteness we consider $n+1$ particles of the
same mass $m$ with a Hamiltonian
$$
H_{tot}=\sum_{A=1}^{n+1}{(\P_A)^2 \over 2m}
+{m\omega^2\over{2(n+1)}}\sum_{A<B=1}^{n+1}
(\R_A-\R_B)^2,
\eqno(1.1)
$$
where $\R_A=(R_{A 1},R_{A 2},R_{A 3})$ and
$\P_A=(P_{A 1},P_{A 2},P_{A 3}),\; A=1,\ldots,n+1,$
are the positions and the momentum operators of the
particles, respectively.

The new features of the WQSs stem from the circumstance that
their position and momentum operators ({\it RP-operators}) do
not in general satisfy the canonical commutation relations (CCRs)
(bellow and throughout $[x,y]=xy-yx$, $\{x,y\}=xy+yx$ ):
$$
[R_{A j},P_{B k}]=i\hbar\delta_{jk}\delta_{A B}
,\quad [R_{A j},R_{B k}]=[P_{A j},P_{B k}]=0, \quad
j,k=1,2,3, \;\; A, B=1,\ldots,n+1. \eqno(1.2)
$$
In certain cases the defining relations are weaker than the CCRs
(the one-dimensional oscillator of Wigner [5], the $osp(1/6)$
oscillators [3]) and therefore the canonical picture appears as a
particular representation of the {\it RP-operators}. In other
cases ([1], the $osp(3/2)$ oscillator [3,4]) the {\it
RP-operators} do not reproduce the canonical picture.  The
present paper is another example of this kind.

The idea  for studing such more general quantum systems belongs
to Wigner [5] (see the discussions in [1-4]), who has generalized
a result of Ehrenfest [6], sometimes refered to as an Ehrenfest
theorem [7]. The latter states  (up to ordering details,
which are important, but will not appear in our considerations)
that in the Heisenberg picture of the quantum mechanics the
Hamiltonian (resp. the Heisenberg) equations are a unique
consequence from the CCRs and the Heisenberg (resp.  Hamiltonian)
equations. In [5] Wigner has proved a stronger statement. He has
shown that for certain interactions the Hamiltonian equations
can be identical to the Heisenberg equations for position and
momentum operators, which do not necessarily satisfy the
canonical commutation relations. Wigner has demonstrated this
on an example of a one-dimensional harmonic oscillator,
studied subsequantly by several authors [8].  This observation is
in the origin of our definition of a Wigner quantum system. The
main point is that the position and the momentum operators ${\bf
R}_A=(R_{A 1},R_{A 2},R_{A 3})$ and ${\bf
P}_A=(P_{A 1},P_{A 2},P_{A 3}),\;
A=1,\ldots,n+1,$ are considered as unknown operators, which
have to be defined in such a way that the Heisenberg equations
$$
{\dot \P_A}=-{i\over\hbar}[\P_A,H_{tot}],\quad
{\dot \R_A}=-{i\over\hbar}[\R_A,H_{tot}] \eqno(1.3)
$$
are identical with the Hamiltonian equations
$$
{\dot
\P_A}=-{m\omega^2\over{n+1}}\sum_{B=1}^{n+1}\big(\R_A-\R_B \big),
\quad {\dot\R_A}={\P_A\over m}. \eqno(1.4)
$$
In addition the {\it RP-operators} have to satisfy other natural
physical requirements. On the first place, they have to be
defined as hermitian operators in a Hilbert space $W$, the state
space of the system. Next, the description should be covariant
with respect to the transformations from the Galilean group $G$.
In particular we have to define the generators of $G$ as
polynomials of $\R_A=(R_{A 1},R_{A 2},R_{A 3})$ and ${\bf
P}_A=(P_{A 1},P_{A 2},P_{A 3}), \; A=1,\ldots,n+1,$ (and to identify
the generators of the space rotation group $SO(3)$ in $G$) so
that $\R_A=(R_{A 1},R_{A 2},R_{A 3})$ and $\P_A =
(P_{A 1},P_{A 2},P_{A 3})$ tranform as vectors.  These  restrictions on
the {\it RP-operators} are in addition to those imposed from the
requirement the Heisenberg equations (1.3) to be identical with
the Hamiltonian equations (1.4).

Our considerations are all of the time in the Heisenberg picture.
We underline that the results depend on the dynamics, since the
solution for $\R_A,\;\P_A$ we are searching for
hold only for the Hamiltonian (1.1).

The paper is organized as follows. In the beginning of Sect. 2 we
state the postulates of a Wigner quantum system. Then accepting
some natural assumptions, which hold in the canonical quantum
mechanics, we separate the centre of mass variables, which are
postulated to be canonical. The rest of the problem is reduced
to a noncanonical $3n$-dimensional Wigner oscillator for the
internal variables, i.e., the Hamiltonian reads
$$
H=\sum_{\alpha=1}^{n} \Big({ {\bf p}_\alpha^2 \over 2m}
+{m\omega^2\over{2}}{\r}_\a^2 \Big). \eqno(1.5)
$$

In Sect. 3 we study one possible solution for ${\bf r}_\alpha,\;
{\bf p}_\alpha ,\quad \alpha =1,\ldots,n$. It is defined in terms
of operators, called creation and annihilation operators (CAOs),
which satisfy certain relations (see (3.8)). In $3.1$ we construct
a class of Fock representations of the CAOs. Each such
representation space is a state space of the oscillator. It is
irreducible and finite-dimensional. The set of all Fock spaces
are labeled with one positive integer $p=1,2,\ldots$. As a result
(Subsect. $3.2$) the spectrum of the Hamiltonian (1.5) is
equidistant and has $min(3n+1,p+1)$ different values. The related
$(n+1)-$particle system has finite space dimensions; the maximal
distance between any two of its constituents is $D=\sqrt{6\hbar
p\over (3n-1)m\omega}$. The internal angular momentum $M$ of the
sistem takes all integer values from 0 to $n$. A particular
feature of the coordinate operators, corresponding to each
particle, is that they do not commute with each other. Therefore,
although the distances between the particles are integrals of
motion, the position of each individual particle cannot be
localized. The kind of noncommutative geometry, obtained in this
way, holds only in a very small volume around the centre of mass.
In a first approximation, namely up to additive terms
proportional to $\sqrt{\hbar}$, the coordinates of all species
coincide with the centre of mass coordinates (see (3.31)). In
$3.3$ we discuss shortly the underlying algebraical structure of
the CAOs. It turns out the creation and the annihilation
operators are odd generators of the orthosymplectic Lie
superalgebra $sl(1/3n)$. Therefore the Fock representations of
the CAOs are in fact representations of this Lie superalgebra.
Section 4 is not directly related to the Wigner quantum systems.
Here we describe shortly the statistics of the creation and the
annihilation operator, called A-superstatistics.  In particular
we formulate the Pauli principle of the A-superstatistics, which
identify it as one of the exclusion statistics of Haldane [9]. It
turns out that the A-superstatistics is very similar to the
statistics of the $g-$ons as introduced by Karabali and Nair
[10]. Some possible applications of the A-superstatistics are also
mentioned. We complete the paper (Sect. 5) trying to justify why
do we interpret the noncanonical operators $\R_\a$ and $\P_\a$ as
position and momentum operators.


\vskip 32pt
\noindent
{\bf 2. Wigner quantum systems}
\bigskip

\noindent
To begin with we give the following  definition of a Wigner
quantum system. A system with a Hamiltonian
$$
H_{tot}=\sum_{k=1}^N {\p_k^2\over 2m_k }+
V(\r_1,\r_2,\ldots,\r_N),
$$
which depends on $6N$ variables $\r_{k}$ and $\p_{k},
\;\;k=1,\ldots,N $, interpreted as
(Cartesian) coordinates and momenta, respectively,
is said to be a Wigner quantum system if the following conditions
hold:

\smallskip
{\bf P1.} The state space $W$ is a Hilbert space. The observables
are Hermitian (selfadjoint) operators in $W$.  The expectation
value $\langle A \rangle$ of the observable $A$ in a state $\phi$
is $\langle A
\rangle = (\phi, A \phi)/(\phi,\phi)$.

\smallskip
{\bf P2.} The Hamiltonian equations and the Heisenberg equations
are identical (as operator equations) in $W$.

\smallskip
{\bf P3.} The description is covariant with respect to the
transformations from the Galilean group.

\smallskip

The postulate {\bf P1} contains the very essence of any quantum
description.  {\bf P2} (Wigner postulate) is weaker than the
requirement the CCRs (1.2) to hold.  Hence the setting is more
general and, consequently, for certain interactions [1-4] the
results differ from the predictions of the canonical theory. In
the general case one has to care about the ordering of the
operators, a problem which does not appear for the Hamiltonian
(1.1). We would not go into discussions of the postulate of
Galilean invariance {\bf P3}. Here the setting is the same as in
the canonical case (see, for instanse [11]). In particular it
insures that the Hamiltonian and the Heisenberg equations do not
prefer any origin in space and time or any direction in space.
The transition probability $|(\psi,\phi)|,\;\psi,\phi \in W $
remains unchanged under the Galilean transfomations of the
states, etc.

We proceed to satisfy {\bf P1-P3} with noncanonical position and
momentum operators. To this end we pass to new variables, which
formally coincide with the Jacoby coordinates and momenta [12],
$$
\eqalign{
&  \R={\sum_{A=1}^{n+1}\R_A \over{n+1}}, \quad
   \P=\sum_{A=1}^{n+1}\P_A,  \cr
& \r_\a={\sum_{\b=1}^\a \R_\b-\a\R_{\a+1}
  \over \sqrt{\a(\a+1)}}, \quad
  \p_\a= {\sum_{\b=1}^\a \P_\b-\a\P_{\a+1}
    \over \sqrt{\a(\a+1)}}, \quad
   \a=1,\ldots,n. \cr
}\eqno(2.1)
$$
\noindent
Then, despite of the fact that $\R, \P, \r_\a, \p_\a$ are unknown
operators, the Hamiltonian $H_{tot}$ splits into a sum of a
centre of mass (CM) Hamiltonian $H_{CM}$ and an internal Hamiltonian
$H$,
$$
H_{tot}=H_{CM}+H, \eqno(2.2)
$$
where
$$
H_{CM}={\P^2\over{2m(n+1)}},\quad
H=\sum_{\alpha=1}^{n} \Big({ {\bf p}_\alpha^2 \over 2m}
+{m\omega^2\over{2}}{\r}_\a^2 \Big). \eqno(2.3)
$$
The Heisenberg equations (1.3) read in terms of (2.1):
$$
{\dot \P}=-{i\over\hbar}[\P,H_{tot}],\quad
{\dot \R}=-{i\over\hbar}[\R,H_{tot}],\quad
{\dot \p}_\a=-{i\over\hbar}[\p_\a,H_{tot}],\quad
{\dot \r}_\a=-{i\over\hbar}[\r_\a,H_{tot}],\quad
\a=1,\ldots,n. \eqno(2.4)
$$
The Hamiltonian equations (1.4) yield:
$$
{\dot \P}=0, \quad {\dot \R}={\P\over m(n+1)},\quad
{\dot \p}_\a=-m\omega^2\r_\a,\quad{\dot \r}_\a={\p_\a\over m},
 \quad \a=1,\ldots,n.
\eqno(2.5)
$$
The problem is to determine operators $\R, \P, \r_\a, \p_\a$ so
that the postulates {\bf P1-P3} hold.  In these variables {\bf
P2} says that eqs. (2.4) have to be equivalent to eqs.  (2.5).
Certainly eqs. (2.4)-(2.5) are satisfied with  canonical
operators (the CCRs bellow follow from (1.2), since the
transformation (2.1) is a canonical one), namely
$$
\eqalignno{
& [R_i,r_{\a j}]=[P_i,r_{\a j}]=[R_i,p_{\a j}]=[P_i,p_{\a j}]=0,
\quad i,j=1,2,3, \quad \a=1,\ldots,n,& (2.6a)\cr
& [R_j,P_k]=i\hbar\delta_{jk},\quad [R_j,R_k]=[P_j,P_k]=0, \quad
j,k=1,2,3, & (2.6b)\cr
& [r_{\a j},p_{\b k}]=i\hbar\delta_{\a \b}\delta_{jk}
,\quad [r_{\a j},r_{\b k}]=[p_{\a j},p_{ \b k}]=0, \quad
j,k=1,2,3, \;\; \a, \b=1,\ldots,n. & (2.6c)\cr
}
$$
We wish to study other, dynamically dependent, solutions. Our
purpose is not to determine all possible operators, satisfying
{\bf P1-P3}. Rather than that we restrict ourselves to
noncanonical solutions only for the internal variables
$\r_\a,\p_\a,\;\,\a=1,\ldots,n.$ In accordance with the canonical
case, we accept

{\it Assumption 1}. The CM variables commute with the internal
variables, i.e., Eqs. (2.6a) hold.

Under this assumption Eqs. (2.4)-(2.5) resolve into two
independent groups, the first one depending only on the CM
position and momentum operators:
$$
\eqalignno{
& {\rm CM\; Hamiltonian\; Eqs.}\quad\quad
  {\dot \P}=0, \quad
  \hskip 18mm {\dot \R}={\P\over m(n+1)},  & (2.7a)  \cr
& {\rm CM\; Heisenberg\; Eqs.}\quad\quad\;\;
  {\dot \P}=-{i\over\hbar}[\P,H_{CM}],\quad
  {\dot \R}=-{i\over\hbar}[\R,H_{CM}].   & (2.7b)  \cr
}
$$
The second group depends only on the internal variables
($\a=1,\ldots,n$):
$$
\eqalignno{
& {\rm Internal\; Hamiltonian\; Eqs.}\quad\quad
  {\dot \p}_\a=-m\omega^2\r_\a,\quad\;\;\;\;{\dot \r}_\a={\p_\a\over m},
     & (2.8a)\cr
& {\rm Internal\; Heisenberg \; Eqs.}\quad\quad\;\;\;
  {\dot \p}_\a=-{i\over\hbar}[\p_\a,H],\quad
  {\dot \r}_\a=-{i\over\hbar}[\r_\a,H].   & (2.8b) \cr
}
$$
With the next assumption we solve equations (2.7).

{\it Assumption 2.} The center of mass coordinates and momenta
are canonical, they satisfy Eqs. (2.6b).

Consequently the centre of mass behaves as a free canonical
point particle with a mass $m(n+1)$.  Thus we are left with the
equations (2.8), which coincide with the Hamiltonian and the
Heisenberg equations of a (noncanonical)  $3n-$dimensional
harmonic oscillator.

Turning to the Galilean covariance, we note that in the canonical
situation the state space $W$ caries a projective representation
of $G$ and of its Lie algebra $g$. It is an exact representation
of the central extension $\hat g$ of $g$ with the generator
of the total mass  of the system. As in the canonical
quantum mechanics, we accept the following identification between
the generators of $g$ and some of the observables of the
$(n+1)-$particle system:

{\it Assumption 3.}
$$
\eqalignno{
1^0\;\; & {\rm The\; angular\; momentum\; operatots}\; \J=\L+\M \; {\rm are
\;generators\;  of\; the\; algebra} \; so(3) & \cr
    & {\rm of\; the \; space \; rotations,}  & (2.9a) \cr
2^0\;\; & H_{tot}=H_{CM}+H \; {\rm is \; a \; generator \; of \; the \;
  translations \; in \; time},   & (2.9b) \cr
3^0\;\; &{\rm  The \; operators \; of \; the \; total \; momentum
   \;{\bf P} \; are \; generators \; of \; the \;
  space \; translations},  & (2.9c) \cr
4^0\;\; &  \K=\mu\R-\P t \; {\rm are \; generators \; of \; the \;
  accelerations}.  & (2.9d) \cr
}
$$
In (2.9) $t$ is the time, $\mu=m(n+1)$ is the mass of the system.
This already means we have chosen a representation of $\hat g$
with a value $\mu$ of the mass operator (which is one of the
Casimir operators).  ${\bf L}=(L_1,L_2,L_3)$ are the operators of
the angular momentum of the centre of mass,
$$
L_i={1\over 2\hbar}\sum_{j,k=1}^3 \varepsilon_{ijk}\{R_j,P_k\},
\eqno(2.10)
$$
which generate also an $so(3)$ algebra, denoted as $so(3)_{CM}$.
${\bf M}=(M_1,M_2,M_3)$ are operators still to be determined.
In the fixed mass representation  the generators of ${\hat g}$ satisfy
the commutation relations $(j,k,l=1,2,3)$ [13]:
$$
\eqalignno{
& [J_j,J_k]=i\varepsilon_{jkl}J_l,\quad \;
   [J_j,P_k]=i\varepsilon_{jkl}P_l,\quad \hskip 8mm
   [J_j,K_k]=i\varepsilon_{jkl}K_l,\quad
   [J_j,H_{tot}]=0, & (2.11a)  \cr
&  [P_j,P_k]=0, \hskip 12mm
   [P_j,K_k]=-i\hbar\delta_{jk}\mu, \hskip 8mm
   [P_j,H_{tot}]=0,  & (2.11b) \cr
&  [K_j,K_k]=0,\hskip 11mm [K_j,H_{tot}]=i\hbar P_j. & (2.11c) \cr
}
$$

The Galilean covariance is to a big extent covered by the
above commutation relations, which have to be satisfied together
with Eqs. (2.8). In particular from (2.11) one concludes that
$\J, \; \P, \; \R$ and $\K$ transform as vectors.  From (2.11)
however does not follow that $\r_\a,\; \p_\a$ are vectors. This
is a problem still to be solved and we will solve it in few
steps.

Observe first of all that the generators of the centre of mass
$\L, \; \P, \K $ and $H_{CM}$ satisfy (2.11) (with $L_j$ instead
of $J_j$). Hence these generators define a (projective)
representation of an algebra, isomorphic to $g$, denoted here  as
$g_{CM}$. This is a representation of a point particle with a mass
$\mu$. The operators $\L$ and $H_{CM}$ generate (a representation
of) the subalgebra $so(3)_{CM}\oplus u(1)_{CM} \subset g_{CM}$.

In the canonical case $\M$ is a vector operator, commuting with
$H$ and both $\M$ and $H$ are in the enveloping algebra of
$\r_\a$ and $\p_\a,\;\; \a=1,\ldots,n$.  More precisely,
$$
M_i=\sum_{\a=1}^n M_{\a i} , \quad
    M_{\a i}={1\over 2\hbar}\sum_{j,k=1}^3
        \varepsilon_{ijk}\{r_{\a j},p_{\a k} \}. \eqno(2.12)
$$
Therefore also here we assume that $\M$ can be expressed in terms
of the internal variables.

{\it Assumption 4.} The components of $\M$ and $H$ are generators
of $so(3)_{int} \oplus u(1)_{int}$.  They are in the enveloping
algebra of the internal position and momentum operators $\r_\a,
\p_\a,\;\;\a=1,\ldots,n$. $\M,\;\r_\a,\;\p_\a$ transform as
vectors with respect to $so(3)_{int}$:

$$
[M_j,M_k]=i\varepsilon_{jkl}M_l, \quad
[M_j,r_{\a k}]=i\varepsilon_{jkl}r_{\a l}, \quad
[M_j,p_{\a k}]=i\varepsilon_{jkl}p_{\a l},\quad \a=1,\ldots,n.
\eqno(2.13)
$$

From Assumption 4 follows that the operators
$$
\J=\L+\M, \;\; \P, \;\; \K=\mu \R-\P t, \;\; H_{tot}=H_{CM}+H
\eqno(2.14)
$$
satisfy Eqs. (2.11). Moreover, the operators
$\R_A, \;\P_A, \; \r_\a, \;\p_\a, \R, \; \P, \; \J, \; \K, \; \;
\L, \; \M \;$ transform as vectors. In particular,
$$
[J_j,A_k]=i\varepsilon_{jkl}A_l \;\; {\rm for \; any}\;\;
{\bf A}\in \{\R_A, \;\P_A, \; \r_\a, \;\p_\a, \R, \; \P, \;
\J, \; \K, \; \; \L, \; \M     \}. \eqno(2.15)
$$
In other words, {\bf P3} is a consequence of Assumption 4. From
the same assumption we conclude that $g_{CM}$ commutes with
$so(3)_{int}\oplus u(1)_{int}$ (with $H$ being a generator of
$u(1)_{int}$).  Therefore the (physical) Galilean algebra $g$ is a
subalgebra of a larger (Lie) algebra $g_{CM}\oplus
so(3)_{int}\oplus u(1)_{int}$. Hence given representation of $g$ is
realized in a state space $W$, which is a tenzor product of the
canonical "free-particle" state space $W_{CM}$ with mass $\mu$
and a module (=representation space) $W_{int}$ of the algebra
$so(3)_{int}\oplus u(1)_{int}$,
$$
W=W_{CM}\otimes W_{int}. \eqno(2.16)
$$
The CM variables $\R, \; \P, \; \L$ are hermitian operators in
$W_{CM}$. Therefore all operators $H_{tot}, \; H_{CM}, \; H, \;
\R_A, \break\hfill
\;\P_A, \; \r_\a, \;\p_\a, \R, \; \P, \; \J, \; \K, \; \;
\L, \; \M \;$ will be Hermitian operators in $W$, if
$\r_\a,\;\p_\a$ and $\M$ are Hermitian operators in $W_{int}$.
Thus condition {\bf P1} holds if (still the unknown operators)
$\r_\a,\;\p_\a$ and $\M$ are hermitian operators in $W_{int}$.

We summarize. The $(n+1)-$particle system with a Hamiltonian
(1.1) is a Wigner quantum system, i.e., the postulates {\bf
P1-P3} hold, if

\smallskip
${\bf P1}_{int}$. The state space $W_{int}$ is a Hilbert space.
The observables (in this case $\r_\a, \; \p_\a, \; \M$ and $H$)
are Hermitian operators in $W_{int}$.

\smallskip
${\bf P2}_{int}$. The internal Hamiltonian equations (2.8a) and
the internal Heisenberg equations (2.8b) are identical (as
operator equations) in $W_{int}$.

\smallskip
${\bf P3}_{int}$ (= {\it Assumption 4}).  The internal
Hamiltonian $H$ and the components of  $\M$ are generators of
$so(3)_{int}\oplus u(1)_{int}$. They are polynomials of the
internal position and momentum operators $\r_\a, \; \p_\a, \;\;
\a=1,\ldots,n$ , so that
$$
[M_j,H]=0,\quad
[M_j,A_k]=i\varepsilon_{jkl}A_l, \quad
A_k \in \{M_i,\; r_{\a i}, \; p_{\a i} |i=1,2,3;\; \a=1,\ldots,n\}.
\eqno(2.17)
$$

The above postulates identify $\r_\a$ and $\p_\a, \;
\a=1,\ldots,n$ as position and momentum operators of a
noncanonical $3n-$dimensional Wigner quantum oscillator.
We proceed to study an example of such an oscillator or, more
precisely, of such oscillators, since the position and the
momentum operators will have several inequivalent
representations.


\vskip 32pt
\noindent
{\bf 3. sl(1/3n) Wigner quantum systems}
\bigskip

\noindent
The problem of constructing  a WQS with a Hamiltonian (1.1) has
been reduced to a problem of building  a Wigner quantum
oscillator, namely a $3n-$dimensional noncanonical oscillator
with a Hamiltonian
$$
H=\sum_{\alpha=1}^{n} \Big({ {\bf p}_\alpha^2 \over 2m}
+{m\omega^2\over{2}}{\r}_\a^2 \Big),\eqno(3.1)
$$
Hamiltonian equations
$$
   {\dot \p}_\a=-m\omega^2\r_\a,\;\;{\dot \r}_\a={\p_\a\over m},
   \quad \a=1,\ldots,n, \eqno(3.2)
$$
and Heisenberg equations
$$
  {\dot \p}_\a=-{i\over\hbar}[\p_\a,H],\quad
  {\dot \r}_\a=-{i\over\hbar}[\r_\a,H],
  \quad \a=1,\ldots,n, \eqno(3.3)
$$
for which the  conditions ${\bf P1}_{int}-{\bf P3}_{int}$ hold.

Eqs. (3.2) and (3.3) are compatible only if
$$
[H,\p_\a]=i\hbar m\omega^2\r_\a,\quad
[H,\r_\a]=-{i\hbar\over m}\p_\a. \eqno(3.4)
$$

In this section we introduce one particular set of Wigner quantum
oscillators, which we call $sl(1/3n)$ oscillators, and
investigate some of the properties of the related $(n+1)-$particle
system. The reason to choose this name is of an algebraic origin
and will be explained in Subsect. {\it 3.3.}

\vskip 22pt

\noindent
{\it 3.1 Satisfying conditions ${\bf P1}_{int}-{\bf P3}_{int}$}
\bigskip

\noindent
Introduce in place of $\r_\a,\;\p_\a$  new unknown operators
$$
a_{\alpha k}^\pm=\sqrt{(3n-1)m \omega \over 4 \hbar} r_{\a k} \pm
{i  \sqrt {(3n-1)\over 4m \omega \hbar}}p_{\a k} , \quad k=1,2,3,
\quad \alpha =1,2,\ldots , n. \eqno(3.5)
$$

\noindent For the sake of convenience we refer to $a_{\alpha
k}^+$ and to $a_{\alpha k}^-$ as to creation and annihilation
operators (CAOs), respectively. These operators should be not
confused with Bose operators.  They  are unknown operators we are
searching for. In terms of these operators the internal
Hamiltonian (3.1) and the compatibility condition (3.4) read
$$
\eqalignno{
& H={\omega \hbar\over{3n-1}}\sum_{\a=1}^n  \sum_{i=1}^3
\{a_{\a i}^+, a_{\a i}^- \}, & (3.6)\cr
& \sum_{\b=1}^n\sum_{j=1}^3  [ \{a_{\b j}^+,a_{\b j}^- \},a_{\a i}^\pm]
=\mp (3n-1)a_{\a i}^\pm , \quad i=1,2,3,
\quad \alpha =1,2,\ldots , n.& (3.7)  \cr
}
$$

As a solution of Eq. (3.7) we choose operators
$a_{\alpha k}^\pm,  \quad k=1,2,3, \quad \alpha =1,2,\ldots , n,$
which satisfy the relations
$$
\eqalignno{
& [\{a_{\a i}^+,a_{\b j}^-\},a_{\g k}^+]=
\delta_{jk}\delta_{\b \g}a_{\a i}^+
-\delta_{ij}\delta_{\a \b}a_{\g k}^+, & (3.8a)\cr
& [\{a_{\a i}^+,a_{\b j}^-\},a_{\g k}^-]=
-\delta_{ik}\delta_{\a \g}a_{\b j}^-
+\delta_{ij}\delta_{\a \b}a_{\g k}^-, & (3.8b)\cr
& \{a_{\a i}^+,a_{\b j}^+\}=
\{a_{\a i}^-,a_{\b j}^-\}=0. & (3.8c) \cr
}
$$

We recall that all considerations are in the Heisenberg picture.
The position and the momentum operators depend on time.
Hence also the CAOs depend on $t$.
Writing the time dependence explicitly, we obtain:
$$
\eqalignno{
& {\rm Hamiltonian \;  equations}\quad\quad
{\dot a}_{\a k}^\pm(t)=\mp i \omega a_{\a k}^\pm(t), & (3.9) \cr
& {\rm Heisenberg \; equations}\quad\quad \;\;\;
{\dot a}_{\a k}^\pm(t)=-{{i \omega }\over{3n-1}}\sum_{\b=1}^n \sum_{j=1}^3
[a_{\a k}^\pm(t),\{a_{\b j}^+(t), a_{\b j}^-(t)\}]. & (3.10) \cr
}
$$
The solution of (3.9) is evident,
$$
a_{\a k}^\pm(t)=exp(\mp i \omega t) a_{\a k}^\pm(0) \eqno(3.11)
$$
and therefore if the defining relations (3.8) hold
at a certain time $t=0$, i.e., for
$a_{\a k}^\pm \equiv a_{\a k}^\pm(0) $,
then they hold as equal time relations for any other time $t$.
From (3.8) it follows also that the Eqs. (3.10) are identical
with Eqs. (3.9). For further references we formulate this result
directly in terms of $\r_\a$ and $\p_\a$.

\smallskip
{\it Conclusion 1.} Within {\it any} representation space $W_{int}$ of
the CAOs (3.8) the Hamiltonian equations (3.2) are identical with
the Heisenberg equations (3.3), i.e., the condition
${\bf P2}_{int}$ holds.
The explicit time dependent solutions of these equations read:
$$
r_{\a k}(t)={\sqrt{\hbar \over {(3n-1)m\omega} }}
(a_{\a k}^+ e^{-i \omega t}+a_{\a k}^- e^{i \omega t}),\quad
p_{\a k}(t)=-i\sqrt{m \omega \hbar \over {3n-1}}
(a_{\a k}^+ e^{-i \omega t}-a_{\a k}^- e^{i \omega t}). \eqno(3.12)
$$

\smallskip
Turning to condition ${\bf P3}_{int}$, we set
$$
M_{\a j}=-i\sum_{k,l=1}^3 \varepsilon_{jkl}\{a_{\a k}^+,a_{\a l}^- \}
=-{{3n-1}\over 2\hbar}\sum_{k,l=1}^3 \varepsilon_{jkl}
\{r_{\a k},p_{\a l} \},
\quad j=1,2,3, \quad \alpha =1,2,\ldots , n. \eqno(3.13)
$$
Then
$$
[H,M_{\a j}]=0,\quad
[M_{\a j},M_{\a k}]=i\sum_{l=1}^3\varepsilon_{jkl}M_{\a l},
\quad j,k,l=1,2,3, \eqno(3.14)
$$
i.e., for each $\alpha =1,2,\ldots , n$ the operators
$\M_\a=(M_{\a 1},M_{\a 2},M_{\a 3})$ are generators of an $so(3)$
algebra, which we denote $so(3)_\a$. Eqs. (3.8)
yield that any two different algebras commute,
$$
[so(3)_\a,so(3)_\b]=0 \quad \forall \a\not= \b=1,\ldots,n.
\eqno(3.15)
$$
It is straightforward to check that the operators $H$ and
$M_i=\sum_{\a=1}^n M_{\a i}$ satisfy Eqs. (2.17). Thus, we have

{\it Conclusion 2.} Within {\it any} representation space $W_{int}$ of
the CAOs (3.8) the operators $H$ and $\M=(M_{ 1},M_{ 2},M_{ 3})$
satisfy the condition ${\bf P3}_{int}$.

It remains to define the (internal) position and the momentum
operators $\r_\a$ and $\p_\a$, corresponding to the CAOs (3.8),
as linear Hermitian operators in a Hilbert space, which will be
the internal state space $W_{int}$. In terms of the creation and
the annihilation operators this means that the Hermitian
conjugate to $a_{\a k}^+$ should be equal to $a_{\a k}^-$, i.e.,
$$
(a_{\a k}^+)^\dagger=a_{\a k}^-.\eqno(3.16)
$$

The CAOs (3.8) have several representations. Here, as in [1],
we consider only representations which are obtained by the usual
Fock space technique. The irreducible Fock representations are
labelled by one non-negative integer $p=1,2,\ldots$, called an
 order of the statistics. To construct them assume that the
corresponding representation space $W(n,p)_{int}$ contains (up to
a multiple) a cyclic vector $|0\ra$, such that
$$
a_{\a i}^-|0\ra=0, \quad
a_{\a i}^-a_{\b j}^+|0\ra=p\delta_{\a \b}\delta_{ij}|0\ra,\quad
i,j=1,2,3, \quad \alpha =1,2,\ldots ,n. \eqno(3.17)
$$
The above relations are enough for reconstructing the
representation space $W(n,p)_{int}$. Let
$$
\Theta\equiv \{\t_{11},\t_{12},\t_{13},\t_{21},\t_{22},\t_{23},\ldots,
\t_{n1},\t_{n2},\t_{n3}  \}.
$$
Since $(a_{\a i}^+)^2=0$,
from (3.17) one derives that the set of all vectors
$$
\eqalignno{
& |p; \Theta \ra \equiv
|p;\t_{11},\t_{12},\t_{13},\t_{21},\t_{22},\t_{23},\ldots,
\t_{n1},\t_{n2},\t_{n3}\ra
=\sqrt{{{(p-\sum_{\a=1}^n  \sum_{i=1}^3
\t_{\a i})!}\over{p!}}}
\prod_{\a=1}^n\prod_{i=1}^3 (a_{\a i}^+)^{\t_{\a i}}|0\ra  & \cr
& \equiv
\sqrt{{{(p-\sum_{\a=1}^n  \sum_{i=1}^3 \t_{\a i})!}\over{p!}}}
(a_{11}^+)^{\t_{11}}(a_{12}^+)^{\t_{12}}(a_{13}^+)^{\t_{13}}
(a_{21}^+)^{\t_{21}}(a_{22}^+)^{\t_{22}}\ldots
(a_{n1}^+)^{\t_{n1}}(a_{n2}^+)^{\t_{n2}}(a_{n3}^+)^{\t_{n3}}|0\ra
& (3.18) \cr
}
$$
with
$$
\t_{\a i}=0,1\quad{\rm and} \quad k\equiv \sum_{\a=1}^n  \sum_{i=1}^3
\t_{\a i} \le p, \eqno(3.19)
$$
constitute an orthonormal basis in  $W(n,p)_{int}$ with respect
to the scalar product, defined in the usual way with "bra" and
"ket" vectors and $\la 0|0 \ra =1$.
We underline that the product of the multiples in
$\prod_{\a=1}^n \prod_{i=1}^3 (a_{\a i}^+)^{\t_{\a i}}$ is ordered
as indicated in (3.18)

Let $|p;\Theta\ra _{\pm \a i} $ be a vector, obtained from
$|p;\Theta\ra$ after a replacement of $\t_{\a i}$ with
$\t_{\a i}\pm 1$. Then the transformation of the basis under the
action of the CAOs read:
$$
\eqalignno{
& a_{\a i}^- |p;\Theta\ra=\t_{\a i}(-1)^{\t_{11}+\ldots +\t_{\a,i-1}}
 \sqrt {p-\sum_{\b=1}^n  \sum_{j=1}^3
 \t_{\b j}+1}\;|p;\Theta\ra _{-\a i}, & (3.20a)\cr
 & a_{\a i}^+ |p;\Theta\ra=(1-\t_{\a i})(-1)^{\t_{11}+\ldots +\t_{\a,i-1}}
 \sqrt {p-\sum_{\b=1}^n  \sum_{j=1}^3
 \t_{\b j}}\;|p;\Theta\ra _{\a i}. & (3.20b)\cr
}
$$
The next conclusion is easily verified.

{\it Conclusion 3.}  The operators $\r_\a, \; \p_\a, \; \M$ and
$H$ are Hermitian operators  within every Hilbert space
$W(n,p)_{int},\;\; p=1,2,\ldots $.

{\it Remark.} The requirement
$\sum_{\a=1}^n  \sum_{i=1}^3 \t_{\a i} \le p$
can be skipped. In such a case one is getting a larger
representation space, which carries an indecomposible
representation of the CAOs. The hermiticity condition (3.16),
however, cannot be satisfied in this larger space. If $p$ is not
a positive integer, (3.16) also cannot be fulfiled in a space with
a positive definite metric.

We have satisfied all requirements of conditions
${\bf P1}_{int}-{\bf P3}_{int}$. Hence within each state space
$W(n,p)_{int}$ the $sl(1/3n)-$oscillator is a Wigner quantum
oscillator and the related $(n+1)-$particle system is a Wigner
quantum system with a state space
$$
W(n,p)=W_{CM}\otimes W(n,p)_{int}. \eqno(3.21)
$$

\vskip 22pt

\noindent
{\it 3.2  Properties of the  $sl(1/3n)$ quantum systems}
\bigskip

\noindent
{\it 3.2.1 Spectrum of the internal Hamiltonian}

\bigskip
\noindent
Note, first of all, that the internal state space $W(n,p)_{int}$
is finite-dimensional.  From (3.6) and (3.20) one concludes that
the internal Hamiltonian $H$ is diagonal in the basis (3.18),
$$
H|p; \Theta \ra ={\omega \hbar\over
{3n-1}}\big(3np-(3n-1)k \big)
|p; \Theta \ra , \eqno(3.22)
$$
where according to (3.19)
$$
k=0,1,2,\ldots,min(3n,p).
\eqno(3.23)
$$
Therefore the internal energy of the  system takes
$min(3n,p)+1$ different values:
$$
E_k={\omega \hbar\over
{3n-1}}\Big(3np-(3n-1)k\Big), \quad k=0,1,2,\ldots,min(3n,p).\eqno(3.24)
$$
As in the canonical oscillator the energy spectrum is
equidistant.  To each energy $E_k$ there correspond ${3n\choose
k}$ (linearly independent) states, namely all basis vectors $|p;
\Theta \ra$ with fixed value of $k$.

\bigskip

\noindent
{\it 3.2.2 Internal angular momentum}

\bigskip
\noindent
The internal state space $W(n,p)_{int}$ carries a reducible
representation of each $so(3)_\a$. The angular momentum of
each oscillating "particle" is either 0 or 1:
$$
\M_\a^2|p; \Theta \ra=0,\;\;
{\rm if}\;\;\t_{\a 1}=\t_{\a 2}=\t_{\a 3} \;\;
{\rm and}\;\;\M_\a^2|p; \Theta \ra=2|p; \Theta \ra \;\;
{\rm otherwise }. \eqno(3.25)
$$
Each basis vector $|p; \Theta \ra$ is an eigenvector of the
square of the intermal angular momentum $\M^2$:
$$
\M^2|p; \Theta \ra = M(M+1)|p; \Theta \ra,\quad
M=1,2,\ldots,n, \eqno(3.26)
$$
i.e., the internal angular momentum of the composite
$(n+1)-$particle system takes all integer values between $0$ and
$n$.  This conclusion holds for any representation of the CAOs.
The multiplicity of each individual value of $\M^2$ depends,
however, on the order of the statistics $p$, namely, on the
representation.

\bigskip

\noindent
{\it 3.2.3 Geometry and space size of the system}
\bigskip
\noindent
Let us consider first the $n-$dimensional Wigner oscillator as
such, independantly of the initial $(n+1)-$particle system.  In
order to avoid confusions, we refer to the constituents of the
oscillator as to oscillating "particles" (or simply "particles").

It is straightforward to check that the square of the radius vector
$\r_\a^2$ of each "particle" commutes with the (internal)
Hamiltonian and, moreover, all operators $\r_\a^2$ commute with
each other,
$$
[H,\r_\a^2]=0, \quad [\r_\a^2,\r_\b^2]=0,
\quad \a,\b=1,\ldots,n. \eqno(3.27)
$$
Hence all operators $\r_\a^2$ can be simultaneously diagonalized.
The basis vectors $|p;\Theta\ra$ are eigenvectors of these
operators:
$$
\r_\a^2|p;\Theta\ra={\hbar\over {(3n-1})m \omega}\Big
(3p-3k+\sum_{i=1}^3 \t_{\a i}  \Big)|p;\Theta\ra,
\quad \a= 1,\ldots,n. \eqno(3.28)
$$
The latter indicates that the "particles" move along spheres
with radiuses
$$
|r_\a|=\sqrt{{\hbar\Big(3p-3k+\sum_{i=1}^3 \t_{\a i}}\Big)
\over {(3n-1})m \omega},\quad
k=0,1,2,\ldots,min(3n,p).\eqno(3.29)
$$
Setting in (3.29)  $k=0$, one obtains the maximal radius. Hence
the spatial size of the oscillator (its diameter) is
$$
d=2\sqrt{3\hbar p\over (3n-1)m\omega}.\eqno(3.30)
$$
The different  "particles" can stay simultaneously on spheres
with different radiuses. The positions of the "particles" on the
spheres, however, cannot be localized, since the coordinate
operators do not commute with each other, $[r_{\a i}, r_{\a
j}]\ne 0, \quad i\ne j=1,2,3$.  The  geometry of the oscillator
is noncommutative.

Let us turn  to the $(n+1)-$particle system. The expressions of
$\R_A$ and $\P_A$ in terms of the Jacoby variables and also in
terms of the CAOs read:
$$
\eqalignno{
\R_A=& \R-\sqrt{{A-1}\over A}\; \r_{A-1} +
 \sum_{\a=A}^n \sqrt{1\over{\a(\a+1)}}\;\r_\a & \cr
 =&\R - {\sqrt{{\hbar (A-1) }\over {(3n-1)A m\omega} }}\;
(\ab_{A-1}^+ +\ab_{A-1}^-)
+ \sum_{\a=A}^n
{\sqrt{{\hbar }\over {(3n-1) \a(\a+1) m\omega} }}\;
(\ab_\a^+ +\ab_\a^-), & (3.31)\cr
\P_A=& {1\over{n+1}}\P-\sqrt{{A-1}\over A}\; \p_{A-1} +
 \sum_{\a=A}^n \sqrt{1\over{\a(\a+1)}}\;\p_\a & \cr
 = & {1\over{n+1}}\P
    +i{\sqrt{{\hbar m\omega (A-1) }\over {(3n-1)A } }}\;
    (\ab_{A-1}^+ -\ab_{A-1}^-)
- i\sum_{\a=A}^n
{\sqrt{{\hbar m\omega }\over {(3n-1) \a(\a+1) } }}\;
(\ab_\a^+ -\ab_\a^-). & (3.32)\cr
}
$$

Therefore also in this case the geometry is noncommutative. A new,
somewhat unexpected feature here is that the distance
operators between the particles do not commute, namely in general
$$
[(\R_A-\R_B)^2,(\R_C-\R_D)^2]\ne 0,\;\; {\rm if}\;\;
(A,B)\ne (C,D).\eqno(3.33)
$$
The only square-distance operator, which is diagonal in the basis
(3.18), is  $(\R_1 - \R_2)^2$. From the general expression (3.31)
we obtain
$$
(\R_1-\R_2)^2=2\r_1^2={2\hbar \over {(3n-1)m\omega}}
\sum_{i=1}^3 \{a_{1i}^+,a_{1i}^- \}.
\eqno(3.34)
$$
Therefore
$$
(\R_1-\R_2)^2|p;\Theta\ra={2\hbar\over {(3n-1})m \omega}\Big
(3p-3k+\sum_{i=1}^3 \t_{1 i}  \Big)|p;\Theta\ra. \eqno(3.35)
$$
Hence the spectrum of
$|\R_1-\R_2|\equiv {\sqrt{ (\R_1-\R_2)^2}}$ reads:
$$
\sqrt{{2\hbar\over {(3n-1})m \omega}\Big
(3p-3k+\sum_{i=1}^3 \t_{1 i}  \Big)}, \quad
k=0,1,2,\ldots,min(3n,p).\eqno(3.36)
$$
In particular the maximal distance $D$ between the
first and the second particles is
$$
D=\sqrt{6\hbar p\over (3n-1)m\omega}.\eqno(3.37)
$$
Since both $H$ and $(\R_1-\R_2)^2$ are diagonal operators, they
commute,
$$
[H,(\R_1-\R_2)^2]=0, \eqno(3.38)
$$
and therefore the distance between the first and the second
particles is preserved in time, it is an integral of motion.

It is natural to expect that the spectrum of
$|\R_A-\R_B|\equiv {\sqrt{ (\R_A-\R_B)^2}}$ for any
$A\ne B\;\; A,B=1,\ldots,n+1$ is the same as those of
$|\R_1-\R_2|$. In particular the maximal distance
between the particles with numbers $A$ and $B$ should be $D$.
Whether this is, however, the case is not so easy to see.
The point is that all our construction is very asymmetrical, it
depends on the way one is numbering the particles. In particular
the Jacoby variables (2.1) and hence also the related CAOs (3.5)
do depend strongly on the fixed numbering.  If one is renumbering the
position and the momentum operators, setting
$$
{\tilde \R}_\a=\R_{\sigma(\a)},\;\;
{\tilde \P}_\a=\P_{\sigma(\a)} \quad
{\rm with}\; \; \sigma\in S_n \; \; {\rm being \; any \; permutation}\;\;
\left(\matrix{1, & 2, & 3, & \ldots, & n \cr
    \sigma(1), & \sigma(2), & \sigma(3),  & \ldots, & \sigma(n)\cr}
        \right),
$$
this will lead to new creation and annihilation operators
${\tilde a}_{\a i}^\pm$ (see (3.5)) and hence in principle to a
new representation space according to (3.17).

In the following we show that the representation (and the
representation space) remains the same, when renumbering the
particles. We  diagonalize also $(\R_A-\R_B)^2$ and show that its
spectrum is the same as of $(\R_1-\R_2)^2$ (see (3.35)).  To this
end we first formulate a simple proposition, which proof is
straightforward.

\noindent
\smallskip
{\it Proposition 1.} Let $S$ be any $n\times n$ symmetric
orthogonal matrix: $S^T=S,\;\; S^TS=1$. Then

$(a)$ The operators
$$
{\tilde a}_{\a i}^\pm=\sum_{\b=1}^n S_{\b \a}{a}_{\b i}^\pm  \eqno(3.39)
$$
\hskip 14mm satisfy (3.8);

$(b)$
$$
H={\omega \hbar\over{3n-1}}\sum_{\a=1}^n  \sum_{i=1}^3
\{a_{\a i}^+, a_{\a i}^- \}=
{\omega \hbar\over{3n-1}}\sum_{\a=1}^n  \sum_{i=1}^3
\{{\tilde a}_{\a i}^+, {\tilde a}_{\a i}^- \}; \eqno(3.40)
$$

$(c)$ If Eqs. (3.17) hold, then
$$
{\tilde a}_{\a i}^-|0\ra=0, \quad
{\tilde a}_{\a i}^-{\tilde a}_{\b j}^+|0\ra=
p\delta_{\a \b}\delta_{ij}|0\ra,\quad
i,j=1,2,3, \quad \alpha =1,2,\ldots ,n. \eqno(3.41)
$$

$(d)$ Let
$P\equiv P(a_{11}^+, a_{12}^+, a_{13}^+, a_{21}^+, a_{22}^+, a_{23}^+,
\ldots,  a_{n1}^+, a_{n2}^+, a_{n3}^+)$ be any polynomial of
the creation operators

\hskip 5mm  and ${\tilde P}\equiv P({\tilde a}_{11}^+, {\tilde a}_{12}^+,
{\tilde a}_{13}^+,
{\tilde a}_{21}^+, {\tilde a}_{22}^+, {\tilde a}_{23}^+,
\ldots,  {\tilde a}_{n1}^+, {\tilde a}_{n2}^+, {\tilde
a}_{n3}^+)$. If $[H,P]=0$, then $[H,{\tilde P}]=0$.

$(e)$ If $\prod_{\a=1}^n \prod_{i=1}^3 (a_{\a i}^+)^{\t_{\a i}}|0\ra$
is an eigenvector of the operator $P$ with an eigenvalue $c(p,\Theta)$,
i.e.,
$$
P\prod_{\a=1}^n \prod_{i=1}^3 (a_{\a i}^+)^{\t_{\a i}}|0\ra=
c(p,\Theta)\prod_{\a=1}^n \prod_{i=1}^3 (a_{\a i}^+)^{\t_{\a i}}|0\ra,
\eqno(3.42)
$$
\hskip 14mm then $\prod_{\a=1}^n \prod_{i=1}^3
({\tilde a}_{\a i}^+)^{\t_{\a i}}|0\ra$
is an eigenvector of ${\tilde P}$, corresponding to the same eigenvalue
$c(p,\Theta)$:
$$
{\tilde P}\prod_{\a=1}^n \prod_{i=1}^3
({\tilde a}_{\a i}^+)^{\t_{\a i}}|0\ra=c(p,\Theta)
\prod_{\a=1}^n \prod_{i=1}^3
({\tilde a}_{\a i}^+)^{\t_{\a i}}|0\ra.\eqno(3.43)
$$

We proceed to find the transformatons of the CAOs under
permutations of the Cartesian coordinates $\R_{\a i}$ and
momenta $\P_{\a i}$. From (2.1) and (3.5) one derives
$$
a_{\a k}^\pm=\sqrt{{(3n-1)m \omega} \over {4\hbar \a(\a+1) }}
\Bigg\{ \sum_{\b=1}^\a (R_{\b k} -\a R_{\a +1, k} \Bigg\}
\pm i\sqrt{{(3n-1)} \over {4m \omega\hbar \a(\a+1) }}
\Bigg\{ \sum_{\b=1}^\a (P_{\b k} -\a P_{\a +1, k} \Bigg\}.
\eqno(3.44)
$$
This relation is preserved if permuting the Cartesian variables:
$$
{\tilde a}_{\a k}^\pm=\sqrt{{(3n-1)m \omega} \over {4\hbar \a(\a+1) }}
\Bigg\{ \sum_{\b=1}^\a ({\tilde R}_{\b k} -\a {\tilde R}_{\a +1, k} \Bigg\}
\pm i\sqrt{{(3n-1)} \over {4m \omega\hbar \a(\a+1) }}
\Bigg\{ \sum_{\b=1}^\a ({\tilde P}_{\b k} -\a {\tilde P}_{\a +1, k} \Bigg\}.
\eqno(3.45)
$$
Consider the simplest permutation, namely a transposition
$$
{\tilde \R}_{A+1}=\R_{A},\;\;{\tilde \P}_{A+1}=\P_{A},\;\;
{\tilde \R}_{A}=\R_{A+1},\;\;{\tilde \P}_{A}=\P_{A+1},\;\;
{\tilde \R}_{C}=\R_{C},\;\;{\tilde \P}_{C}=\P_{C},\;\;{\rm if}\;\;
C\ne A, A+1.\eqno(3.46)
$$
Replacing in (3.45) ${\tilde R}_{C k}$ and  ${\tilde P}_{C k}$ with
${R}_{C k}$ and ${ P}_{C k}$, $C=1,\ldots,n+1$, and expressing the
latter  through the CAOs from (3.31) and (3.32) we obtain:
$$
{\tilde a}_{\a k}^\pm=\sum_{\b=1}^n (s_{A+1,A})_{\b \a}a_{\b k}^\pm,
\eqno(3.47)
$$
where $s_{A+1,A}$ is $n\times n$ matrix with the following
nonzero matrix elements:
$$
\eqalign{
& (s_{A+1,A})_{A-1,A-1}=-(s_{A+1,A})_{A,A}={1\over A},\;\;
  (s_{A+1,A})_{\a,\a}=1,\;\;\a\ne A-1, A;  \cr
& (s_{A+1,A})_{A-1,A}=(s_{A+1,A})_{A,A-1}=
  {\sqrt{A^2-1}\over A}.  \cr
}\eqno(3.48)
$$
Eq. (3.47) gives the transformation of the
CAOs, corresponding to the transposition (3.46).

For any $A<B=1,\ldots,n+1$ set
$$
S_{A,B}=s_{A,A-1}s_{A-1,A-2}s_{A-2,A-3}\ldots s_{2,1}
         s_{B,B-1}s_{B-1,B-2}s_{B-2,B-3}\ldots s_{3,2}.\eqno(3.49)
$$
The above matrix leads to a transformation of the CAOs
$$
{\tilde a}_{\a k}^\pm=\sum_{\b=1}^n (S_{A,B})_{\b \a}a_{\b k}^\pm,
\eqno(3.50)
$$
corresponding to a transposition $2\leftrightarrow B$, followed by
$1 \leftrightarrow A$ of the Cartesian variables. Then from
(3.34) and (3.50) we derive:
$$
(\R_A-\R_B)^2={2\hbar \over {(3n-1)m\omega}}
\sum_{i=1}^3 \{{\tilde a}_{1i}^+,{\tilde a}_{1i}^- \}.
\eqno(3.51)
$$
The matrix $S_{A,B}$ satisfies the requirements of proposition 1:
it is symmetric and $S_{A,B}^TS_{A,B}=1$. Therefore the operators
(3.50) satisfy $(b)$ and $(c)$ of proposition 1. Consequently,
$(b)$, $(3.34)$ and (3.51) yield:
$$
[H,(\R_A-\R_B)^2]=0, \eqno(3.52)
$$
whereas from $(c)$ one concludes that the Fock space
corresponding to ${\tilde a}_{\a k}^\pm$ is the same as of
${a}_{\a k}^\pm$.
Writting (3.35) in the form
$$
{2\hbar \over {(3n-1)m\omega}}
\sum_{i=1}^3 \{a_{1i}^+,a_{1i}^- \}
\prod_{\a=1}^n\prod_{i=1}^3 (a_{\a i}^+)^{\t_{\a i}}|0\ra
={2\hbar\over {(3n-1})m \omega}\Big
(3p-3k+\sum_{i=1}^3 \t_{1 i}  \Big)
\prod_{\a=1}^n\prod_{i=1}^3 (a_{\a i}^+)^{\t_{\a i}}|0\ra,
\eqno(3.53)
$$
and applying (e) of proposition 1, we have
$$
{2\hbar \over {(3n-1)m\omega}}
\sum_{i=1}^3 \{{\tilde a}_{1i}^+,{\tilde a}_{1i}^- \}
\prod_{\a=1}^n\prod_{i=1}^3 ({\tilde a}_{\a i}^+)^{\t_{\a i}}|0\ra
={2\hbar\over {(3n-1})m \omega}\Big
(3p-3k+\sum_{i=1}^3 \t_{1 i}  \Big)
\prod_{\a=1}^n\prod_{i=1}^3 ({\tilde a}_{\a i}^+)^{\t_{\a i}}|0\ra,
$$
i.e. (see (3.51),
$$
(\R_A-\R_B)^2
\prod_{\a=1}^n\prod_{i=1}^3 ({\tilde a}_{\a i}^+)^{\t_{\a i}}|0\ra
={2\hbar\over {(3n-1})m \omega}\Big
(3p-3k+\sum_{i=1}^3 \t_{1 i}  \Big)
\prod_{\a=1}^n\prod_{i=1}^3 ({\tilde a}_{\a i}^+)^{\t_{\a i}}|0\ra.
\eqno(3.54)
$$
Thus $\prod_{\a=1}^n\prod_{i=1}^3
({\tilde a}_{\a i}^+)^{\t_{\a i}}|0\ra$
are the eigenvectors of $(\R_A-\R_B)^2$ and hence the spectrum of
$|\R_A-\R_B|$ is (3.36), i.e.,
the same as of $|\R_1-\R_2|$.

The conclusion is that the distances between the particles are
quantized. The maximal distance is $D$ as given with (3.37).
Hence the space size of the composite system is $D$.  The system
exibits a nuclear kind structure: the $n+1-$particles  move in a
small volume around the centre of mass. Since the coordinates do
not commute, the particles are smeared with certain probability
within the volume.  For any two particles with, say, numbers $A$
and $B$ one can always diagonalize $|\R_A-\R_B|$ simultaneously
with the Hamiltonian, i.e., the distance is preserved in time. It
is not possible however to diagonalize simultaneously all
distance operators, since (see (3.33)) they do not commute.

Let us try to analyze the reason and the amount of the
noncommutativity of the coordinates. Since these
operators act in the space
$W(n,p)=W_{CM}\otimes W(n,p)_{int}$,
a more rigorous way to write (3.31) is
$$
\R_A= \R\otimes {\bf 1}+{\bf 1}\otimes
\Big\{ -\sqrt{{A-1}\over A}\; \r_{A-1} +
 \sum_{\a=A}^n \sqrt{1\over{\a(\a+1)}}\;\r_\a\Big\}, \eqno(3.55)
$$
where {\bf 1} is the unity operator (in the corresponding spase).
In the canonical quantum mechanic all operators, having a
classical analogue, and in particular the coordinates of the
$A^{th}$ particle $\R_A$ are operators only in $W_{CM}$, i.e.,
$\R_A=\R\otimes {\bf 1}$. The only operator acting nontrivially
in $W(n,p)_{int}$ is the spin operator $\M$. In our case, due to
the second term in (3.55), also the coordinate operators
transform $W(n,p)_{int}$.  The second terms are small, they are
proportional to $\sqrt{\hbar}$ (see (3.31)), and therefore in a
first approximation can be neglacted. If so, then the coordinate
operators of all particles $\R_A$ coincide with the centre of
mass coordinates $\R$ and the composite system behaves as a
canonical point particle with mass $\mu=m(n+1)$. The terms
$\Big\{ -\sqrt{{A-1}\over A}\; \r_{A-1} + \sum_{\a=A}^n
\sqrt{1\over{\a(\a+1)}}\;\r_\a\Big\}$ in (3.55) split the point
particle into $n+1$ "pieces", which move in a volume with linear
dimension $D$ around the centre of mass. Only within this small
volume the coordinates do not commute. To check however this
"experimentaly" is nontrivial, since it is imposible to isolate
one of the particles, taking it away from the centre of mass.

\vskip 22pt

\noindent
{\it 3.3  A short insight into the algebraic structure}
\bigskip

\noindent
In the present subsection (see also [1]) we discuss shortly the
underlying Lie superalgebraical structure of the creation and the
annihilation operators (3.8). The presentation is independent from the
other part of the paper.

As we have already indicated, any $3n$ pairs of canonical
position and momentum operators, namely operators with relations
(2.6c), provide the simplest solution of ${\bf P1}_{int}
-{\bf P3}_{int}$. It is not so well known that these operators can
be considered as odd generators of a Lie superalgebra. The simplest
way to show this is to pass to the related Bose creation and
annihilation operators:
$$
b_{\a k}^\pm=\sqrt{m \omega \over 2 \hbar} r_{\a k} \mp
{i \over \sqrt {2m \omega \hbar}}p_{\a k},
\quad \a=1,\ldots,n,\;\;k=1,2,3. \eqno(3.56)
$$
It is straightforward to show that the Bose CAOs give one particular
representation, the infinite-dimensional Fock representation,
of the relations
$$
[ \{B_{\a i}^\xi,B_{\b j}^\eta \},B_{\gamma k}^\epsilon]=
\delta_{\a \gamma}\delta_{ik}(\epsilon -\xi)B_{\b j}^\eta
+ \delta_{\b \gamma}\delta_{jk}(\epsilon - \eta)B_{\a i}^\xi, \;
i,j,k=1,2,3,\;\a,\b,\gamma=1,\ldots,n, \;
\xi, \eta, \epsilon =\pm \;{\rm or}\; \pm 1.
\eqno(3.57)
$$

Any set of operators $B_{\a i}^\pm$ with relations (3.57)
generate a Lie superalgebra (LS) [14]. It turns out [15] this is
the orthosymplectic LS $osp(1/6n)$. The operators $B_{\a i}^\pm$
are its odd generators, whereas all anticommutators $\{B_{\a
i}^\xi,B_{\b j}^\eta \}$ span the even subalgebra $sp(6n)$. In
the terminology of Kac [16] $osp(1/6n)$ is one of the basic Lie
superalgebras. Let us  mention that in the quantum field
theory the operators $B_{\a i}^\pm$ are known as para-Bose
operators. They were introduced by Green as a possible
generalization of the statistics of integer spin fields [17].

The creation and the annihilation operators (3.8) generate also a
basic Lie superalgebra [1]. Although its relations are similar to
(3.57), the algebra is very different. In this case it is the
special linear Lie superalgebra $sl(1/3n)$. Its odd generators are
the CAOs; all anticommutators $\{a_{\a i}^+,a_{\b j}^-\}$
span the even subalgebra, which is the Lie algebra $gl(3n)$.
This is the reason to call the $(n+1)$-particle system with
CAOs (3.8) an $sl(1/3n)$ Wigner quantum system. Resently Okubo
has shown that the CAOs (3.8) can be viewed also as generators
of a Lie supertriple system [18].

Any representation of the CAOs (3.8) defines a representation of
$sl(1/3n)$ and vice versa. Therefore the question to determine
the representations of the CAOs (3.8) is equivalent to the
problem to construct the representations of $sl(1/3n)$. The
hermiticity condition (3.16) defines an antilinear antiinvolution
on $sl(1/3n)$. By definititon the representations in Hilbet
spaces, which satisfy (3.16) are called unitary representations
(with respect to this antiinvolution). It turns out all such
representations are finite-dimensional. They were explicitly
constructed in [19] and are labelled with $3n$ numbers, the
coordinates of the highest weight. Therefore the Fock
representations, considered here, give a small part of all
representations, for which the conditions ${\bf P1}_{int} -{\bf
P3}_{int}$ can be satisfied.

Elsewhere we will consider Wigner quantum systems with CAOs
generating another basic LS, namely $sl(n/3)$. The LSs
$sl(n/3)$ and $sl(1/6n)$ belong to the class {\bf A} in the
classification of Kac [16], whereas the algebras $osp(1/6n)$ are
from the class {\bf B}. There are two more infinite classes {\bf
C} and {\bf D} of basic LSs. It will be interesting to see
whether one can introduce CAOs, corresponding to some of them.
Certainly one needs not to restrict to solutions, which generate
simple LSs. The oscillator conditions ${\bf P1}_{int} -{\bf
P3}_{int}$ can be satisfied with semisimple LSs and in particular
with direct sums of LSs as for instance
$$
\oplus_{i=1}^n sl(1/3)\;\;{\rm or}\;\;\oplus_{i=1}^n osp(3/2).
\eqno(3.58)
$$
This possibility will be a subject of future considerations.


\vskip 32pt
\noindent
{\bf 4. Statistics of the creation and the annihilation operators}
\bigskip

\noindent
Here we discuss shortly the statistics, corresponnding to the
algebra of the operators (3.8) and compare it with the very
similar statistics of the $g-$ons [10].

To this end we interpret $a_{\a i}^+$ (resp.  $a_{\a i}^-$)  as
an operator creating (resp.  annihilating) a particle in a
(one-particle) state (= orbital) $(\a i)$. Then the Pauli
principle of the statistics, corresponding to the CAOs (3.8) says
that on every orbital there cannot be more that one particle
(Fermi-kind property, following from (3.8c)). In addition to
this, however, it requires that no more than $p$ orbitals can be
simultaneously occupied. The latter is due to the requirement
(3.19), namely $\sum_{\a=1}^n\sum_{i=1}^3 \theta_{\a i}\le p$.
If, for instance, certain $p$ orbitals are occupied, then the
possible change $\Delta \t_{\b j}$ of the occupation numbers of
any other orbital is zero, $\Delta \t_{\b j}=0$. Therefore the
$A$-superstatistics is among the exclusion statistics, introduced
by Haldane [9].  In fact it is very similar to  the
statistics of the $g-$ons as defined by Karabali and Nair.  We
refer to it as to Karabali-Nair statistics (KN-statistics). The
latter goes beyond the thermodynamic formulation, attempting a
microscopic desctiption of the many-body state space, generated
out of a vacuum vector with polynomials of creation and
annihilation operators $a_{\a i}^\pm$ (we keep close to our
notation). In order to compare it with the statistics of the CAOs
(3.8), we recall the main assumptions of the KN-statistics.

(1) If $|\Theta\ra\equiv
|\t_{11},\t_{12},\t_{13},\t_{21},\t_{22},\t_{23},\ldots,
\t_{n1},\t_{n2},\t_{n3}\ra$ is a state with $\t_{\a i}$ particles
on the orbital $(\a i)$, then
$$
a_{\a i}^\pm |\Theta\ra= c^\pm(\Theta)|\Theta\ra _{\pm\a i}
\equiv |\t_{11},\t_{12},\ldots,\t_{\a i}\pm 1,\ldots
\t_{n2},\t_{n3}\ra , \eqno(4.1)
$$
where $c^\pm(\Theta)$ are constants, depending on the statistics.

(2) There exists a number operator $N_{\a i}$ of the particles on
the orbital $(\a i)$, which is a function of
$a_{\a i}^+ a_{\a i}^-$, so that
$$
[N_{\a i},a_{\b j}^\pm ]=
\pm\delta_{\a \b}\delta_{i j}a_{\b j}^\pm. \eqno(4.2)
$$
Therefore
$$
N_{\a i}|p,\Theta\ra=\t_{a i}|p,\Theta\ra . \eqno(4.3)
$$

(3) For any numbers $c_{\a i}$ there exists an integer $m$,
so that
$$
\Big( \sum_{\a=1}^n\sum_{i=1}^3 c_{\a i}a_{\a i}^\pm \Big)^{m+1}=0.
\eqno(4.4)
$$

(4) The CAOs satisfy the relations
$a_{\a i}^- a_{\b j}^-=R_{\a i,\b j}a_{\b j}^- a_{\a i}^- $,
where $R_{\a i,\b j}$ are numbers.

(5) The CAOs satisfy in addition the relation
$[a_{\a i}^+a_{\a i}^-,a_{\b j}^-]=0$.

Clearly the main properties of the Fock representation of the
CAOs (3.8) in $W(p,n)$ are very similar to the KN-statistics.
Assumption (1) is the same as (3.20). The number operator
reads
$$
N_{\a i}={p\over{3n-1}}+\{a_{\a i}^+,a_{\a i}^-\}
-{1\over{3n-1}}\sum_{\b=1}^n\sum_{i=1}^3\{a_{\b j}^+,a_{\b j}^-\} .
\eqno(4.5)
$$
Therefore (2) also holds, but $N_{\a i}$ is not a function only of
$a_{\a i}^+ a_{\a i}^-$, but of all creation and annihilation
operators.  Assumption (3) holds in our case for $m=p$ and (4) is
fulfiled with $R_{\a i,\b j}=-1$. The assumption (5)
of the KN-statistics is not satisfied in our case.

Finally, we mention that the creation and the annihilation
operators (3.8) (with $n=\infty$) were introduced for the first
time in quantum field theory as a possible generalization of the
statistics of the tenzor fields [20]. In that case they generate
the infinite-dimensional Lie superalgebra $sl(1/\infty)$. The
corresponding statistics was called $A-$superstatistics.
Recently a representation of the $A-$superstatistics,
corresponding to $p=1$ and called ortho-Fermi statistics was
independently proposed by Mishra and Rajasekaran [21].


\vskip 32pt
\noindent
{\bf 5. Concluding remarks}
\bigskip

\noindent
The most difficult question to answer in relation to the present
approach is why do we interpret the noncanonical operators
${\R}_\a$ and ${\P}_\a$ as coordinates and momenta. A rigorous
proof to this question we cannot give.  There exists however no
proof why the CCRs should necessary hold.  This has been noted
already by Wigner [5]. All main quantum postulates are satisfied
by any WQS. A criterion for accepting or rejecting a given WQS
have to be its predictions and finally the experiment. In this
respect some of the predictions of the WQSs are of interest.
Quite new, nonconventional feature of the $sl(1/3n)-$quantum
system, for instance, is its finite size. The particles move in a
small volume around the centre of mass. In a first approximation
(see (3.31)), neglecting the terms proportional to $\sqrt \hbar$,
${\R}_\a$ and ${\P}_\a$ are canonical, they coincide with $\R$
and $\P$. The noncommutativity of the coordinates and, more
generally, the deviation from the CCRs, is due to  small,
proportional to $\sqrt \hbar$, terms added to the CM coordinates
and momenta. As a result a point particle of mass $\mu=m(n+1)$
splits into $n+1$ "pieces" with mass $m$. Only those small,
proportional to $\sqrt \hbar$, coordinates of the "pieces" with
respect to the centre of mass are noncommutative.  In this way
the canonical point particle is "dressed" with internal structure
and it is this "dressing", which is noncanonical. In the limit
$\hbar \rightarrow 0$ the structure disappears; all $n+1$
"pieces" fall onto the centre of mass. The composite system
becomes again a free point particle.  It seems to us that such a
picture deserves an attention.  After all it is unclear so far
whether the protons and the neutrons within a nucleus or, say,
the constituents of a hadron, the quarks, are canonical.

In answering the above question we could have been also more
formal. Nowadays, following the ideas of Connes [22], a lot of
work is done in the field of the noncommutative geometry. The
quantum groups and the related to them deformed oscillators (see
[23] for a list of references) provide other examples of
noncanonical quantum systems.

\vskip 24pt
\noindent
{\bf Acknowledgments}

\bigskip
\noindent
We are grateful to Prof. Randjbar-Daemi for the invitation and
for the kind hospitality at the High Energy Section of ICTP. The
work was supported by the Grant $\Phi-416$ of the Bulgarian
Foundation for Scientific Research.

\vskip 24pt
\noindent
{\bf References}

\vskip 12pt
\settabs \+  [11] & I. Patera, T. D. Palev, Theoretical
   interpretation of the experiments on the elastic \cr

\+ [1] & Palev T D 1982  {\it J. Math. Phys.} {\bf 23} 1778;
         1982 {\it Czech. J. Phys. B} {\bf 3} 680 \cr

\+ [2] & Kamupingene A H, Palev T D and Tsaneva S P 1986
         {\it J. Math. Phys.} {\bf 27} 2067 \cr

\+ [3] & Palev T D and Stoilova N I 1994 {\it J. Phys. A: Math.
         Gen.} {\bf 27} 977  \cr

\+ [4] & Palev T D and Stoilova N I 1994 {\it J. Phys. A: Math.
         Gen.} {\bf 27} 7387  \cr

\+ [5] & Wigner E P 1950 {\it Phys. Rev.} {\bf 77} 711 \cr

\+ [6] & Ehrenfest P, Z. Phys. 4 (1927) 455 \cr

\+ [7] & Messia A 1964 {\it Quantum Mechanics} (North-Holland
         Publ. Co. Amsterdam) \cr

\+ [8] & Ohnuki Y and Kamefuchi S 1978 {\it J. Math. Phys.} {\bf 19} 67;
         1979 {\bf Z. Phys. C} {\bf 2} 367\cr

\+     & Okubo S 1980{\it  Phys. Rev.} {\bf D 22} 919\cr

\+     & Mukunda N, Sudarshan E C G, Sharma J K and Mehta C L,
         1980 {\it J. Math. Phys.} {\bf 21} 2386 \cr

\+     & Ohnuki and Watanabe S, 1992 {\it J. Math. Phys.}
         {\bf 33} 3653.\cr

\+ [9] & Haldane F D M 1991 {\it Phys. Rev. Lett.} {\bf 67} 937\cr

\+ [10] & Karabali D and Nair V P 1995 {\it Nucl. Phys.} {\bf B 438}
          551 \cr

\+ [11] & O'Raifeartaigh L 1970 {\it Lect. Notes in Phys.}
          {\bf 6} 144 (Editor Bargman V, Springer      \cr

\+ [12] & Moshinsky M 1969 {\it The Harmonic Oscillator in Modern
          Physics: From Atoms to Quarks}  \cr
\+      & (Gordon and Breach Sci. Publ.,
           Berlin-Heidelberg-New York) \cr

\+ [13] & Hamermesh M 1960 {\it Ann. of Phys.} {\bf 9} 518\cr

\+ [14] & Omote M, Ohnuki Y and Kamefuchi S 1976
        {\it Prog. Theor. Phys.} {\bf 56} 1948 \cr

\+ [15] & Ganchev A Ch and Palev T D 1980
          {\it J. Math. Phys.} {\bf 21} 797 \cr

\+ [16] & Kac V G  1978 {\it Lect. Notes Math.} {\bf 676} 597 \cr

\+ [17] & Green H S 1953 {\it Phys. Rev.} {\bf 90} 270 \cr

\+ [18] & Okubo S 1994 {\it J. Math. Phys.} {\bf 35} 2785 \cr

\+ [19] & Palev T D     1987 {\it Funct. Anal. Appl.} {\bf 21} 245
         (English transl.); 1989 {\it J. Math. Phys.} {\bf 30}
                 1433  \cr

\+ [20] & Palev T D 1978 {\it Communication JINR} {\it E2-11942};
          1978 {\it Preprint JINR} {\bf E2-11929, P2-11943};\cr
\+      & 1979 {\it Czech. J. Phys.} {\bf B29} 91;
          1980  {\it J. Math. Phys.} {\bf 21} 2560 \cr

\+ [21] & Mishra A K and Rajasekaran G 1991 {\it Pramana-J. Phys.}
          {\bf 36} 537\cr

\+       & Mishra A K and Rajasekaran G 1992 {\it Pramana-J. Phys.}
          {\bf 38} L411 \cr

\+ [22] & Connes A and Lott J 1991 {\it Nucl. Phys (Proc.
          Suppl.)} {\bf 18B} 29 \cr

\+ [23] & Chari V and Pressley A 1994 {\it A Guide to Quantum
          Groups } (Cambridge: Cambridge University Press)\cr

\end